\documentclass[prl,twocolumn,superscriptaddress,showpacs]{revtex4}
\usepackage{amsmath,amssymb,graphicx}

\begin{document}

\title{Stimulated emission of terahertz radiation by semiconductor
microcavities}
\author{K.V. Kavokin}
\affiliation{A.F. Ioffe Physical-Technical Institute, 26, Politekhnicheskaya, 194021,
St-Petersburg, Russia}
\affiliation{International Institute of Physics, Av. Odilon Gomes de Lima, 1772, Capim
Macio, 59078-400, Natal, Brazil}
\author{M.A. Kaliteevski}
\affiliation{Department of Physics, Durham University, Durham DH1 3LE, United Kingdom}
\author{R. A. Abram}
\affiliation{Department of Physics, Durham University, Durham DH1 3LE, United Kingdom}
\author{A.V. Kavokin}
\affiliation{Physics and Astronomy School, University of Southampton, Highfield,
Southampton SO17 1BJ, United Kingdom}
\author{S. Sharkova}
\affiliation{Academic University, 8/3, Khlopin str., 195220 St-Petersburg, Russia}
\author{I. A. Shelykh}
\affiliation{Science Institute, University of Iceland, Dunhagi-3, IS-107, Reykjavik,
Iceland}
\affiliation{International Institute of Physics, Av. Odilon Gomes de Lima, 1772, Capim
Macio, 59078-400, Natal, Brazil}
\affiliation{Academic University, 8/3, Khlopin str., 195220 St-Petersburg, Russia}
\date{\today }

\begin{abstract}
We show that planar semiconductor microcavities in the strong coupling
regime can be used as sources of stimulated terahertz (THz) radiation.
Emitted THz photons would have a frequency of the spliting of the cavity
polariton modes. The optical transition between upper and lower polariton
branches is allowed due to mixing of the upper polariton state with one of
excited exciton states. This transition is stimulated in the polariton laser
regime.
\end{abstract}

\maketitle

The search for effective terahertz (THz) radiation sources and detectors is
one of the important trends of modern applied physics \cite{Dragoman,Davies}%
. None of the existing THz emitters universally satisfies the application
requirements. For example, the emitters based on nonlinear-optic frequency
down-conversion are bulky, expensive, power consuming. Various semiconductor
\cite{Hu} and carbon-based \cite{Portnoi,Wright} devices based upon
intraband optical transitions are compact but have a limited wavelength
adjustment range and have a low quantum efficiency. Among the factors which
limit efficiency of semiconductor THz sources are is the short life-time of
involved electronic states (typically, fractions of a nanosecond) compared
to a long typical time of spontaneous emission of a THz photon (typically
milliseconds). The attempted ways to improve this ratio is the use of the
Purcell effect\cite{Purcell,Gerard,Todorov} in the THz cavities or the
cascade effect in quantum cascade lasers \cite{Faist} (QCL). Nevertheless,
till now the QCL in the spectral region about 1THz remain costly,
short-living and still show the quantum efficiency of less than 1\%. Here we
explore the possibility of generating THz radiation by using the stimulated
transition between bosonic quantum states formed due to strong light-matter
coupling in semiconductor microcavities, namely, exciton polaritons. The
quantum efficiency of this source is governed by the population of the final
polariton state, which may be tuned in large limits by means of the optical
pumping.

In the strong coupling regime in a microcavity \cite{KavokinBook} the
dispersion of the exciton-polaritons is composed by two bands both having
minima at zero in-plane wave vector $\mathbf{k}$. At $\mathbf{k}=0$, the
energy splitting between the two branches approximately equals $\hbar \Omega
_{R}$, where $\Omega _{R}$ is the optical Rabi frequency, the measure of the
light-matter coupling strength in a microcavity. Typically, $\hbar \Omega
_{R}$ is of the order of several meV, which makes this system attractive for
THz applications. Stimulated scattering of exciton polaritons into the
lowest energy state leads to so-called polariton lasing, recently observed
in GaAs, CdTe and GaN based microcavities \cite{Bloch, Levrat}. If the
scattering from upper to lower polariton branch was accompanied by emisson
of THz photons, polariton lasers would emit THz radiation, and this emission
would be stimulated by the population of the lowest energy polariton state.
However, at the first glance, this process is forbidden: an optical dipole
operator cannot directly couple the polariton states formed by the same
exciton state.

This obstacle can be overcome if one of the polariton states of
interest is mixed with an exciton state of a different parity, say,
the e1hh2 exciton state formed by an electron at the lowest energy
level in a quantum well (QW) and a heavy hole at the second energy
level in the QW. This state is typically a few meV above the exciton
ground state e1hh1. By a proper choice of the QW width and
exciton-photon detuning in the microcavity it can be brought into
resonance with the lowest energy upper polariton state. Being
resonant, the two states can be easily hybridised by any weak
perturbation, such as e.g. a built in or external electric field.
The optical transition between such a hybridised state and the
lowest e1hh1 exciton polariton state is allowed.

We consider a model system shown at Fig. 1. It consists of a planar
microcavity in the strong coupling regime, placed inside a planar THz
cavity, which additionally enhances the rate of the emission of the THz
photons.Together with the waveguiding effect of microcavity structure,
adding the lateral THz cavity would realize an effective 3d confinement of
the THz mode, giving rise to enhancement of spontaneous emission rate via
Purcell effect\cite{Purcell,Gerard,Todorov}.
\begin{figure}[tbp]
\begin{center}
\includegraphics[width=0.8\linewidth]{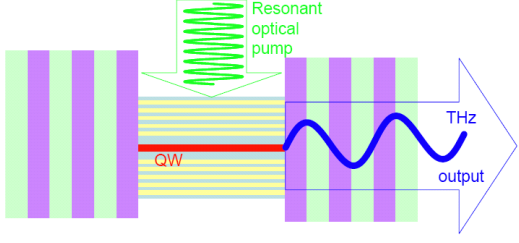}
\end{center}
\caption{Scheme of the polariton terahertz emitter (shown not to scale). A\
planar microcavity is embedded in the lateral terahertz cavity}
\label{fig1}
\end{figure}
The eigenstates of the system can be obtained from diaginalization of the
Hamiltonian matrix, written in the basis of the bright exciton, dark exciton
and cavity photon (see the scheme in Fig. 2).
\begin{equation}
\widehat{H}=\left(
\begin{array}{ccc}
E_{1,ex} & \delta _{\epsilon }/2 & \hbar \Omega _{R}/2 \\
\delta _{\epsilon }/2 & E_{2,ex} & 0 \\
\hbar \Omega _{R}/2 & 0 & E_{ph}%
\end{array}%
\right)  \label{Hamiltonian}
\end{equation}
where $E_{1,ex},E_{2,ex}$ and $E_{ph}$ are the energies of bright and dark
excitons and the cavity photon mode respectively, $\delta _{\epsilon }$ is a
parameter describing the mixture between bright and dark exciton states due
to the electric field $\epsilon $ normal to the QW plane, $\delta _{\epsilon
}\approx e\varepsilon \int_{-\infty }^{+\infty }u_{hh1}(z)zu_{hh2}(z)dz$
with $u_{hh1}(z)$ and $u_{hh2}(z)$ being the confined wavefunctions of the
ground and first excited states of the heavy hole in a QW. One can easily
estimate that in order to have $\delta _{\epsilon }$ larger than the upper
polariton linewidth, which is typically of the order of 0.1 meV, it is
enough to apply an electric (piezoelectric) field of 3 kV/cm in a 10 nm-wide
quantum well. The diagonalization of the Hamiltonian \ref{Hamiltonian}
allows finding the eigenenergies and eigenstates of the system. In the case
of $\hbar \Omega _{R}$, $|E_{ex,2}-E_{1}-V_{R}/2|\ll V_{R}$ and zero
detuning between bright exciton and photon modes ($E_{1,ex}=E_{ph}=E_{1})$
the latter read:
\begin{eqnarray}
|L\rangle &\approx &\frac{1}{\sqrt{2}}\left( |1,ex\rangle +|ph\rangle
\right) , \\
|U+\rangle &\approx &\frac{1}{\sqrt{1+b_{-}^{2}}}\left[ \frac{1}{\sqrt{2}}%
\left( |1,ex\rangle -|ph\rangle \right) +b_{-}|2,ex\rangle \right] , \\
|U-\rangle &\approx &\frac{1}{\sqrt{1+b_{+}^{2}}}\left[ \frac{1}{\sqrt{2}}%
\left( |1,ex\rangle -|ph\rangle \right) +b_{+}|2,ex\rangle \right] .
\end{eqnarray}%
with
\begin{equation}
b_{\pm }=\delta _{\epsilon }^{-1}\left[ E_{2}-V_{R}/2-E_{1}\pm \sqrt{%
(E_{1}+V_{R}/2-E_{2})^{2}+\delta _{\epsilon }^{2}}\right] .
\end{equation}%
The indices $L,U+,U-$ correspond to the lower polariton branch and two upper
branches formed due to the mixture between upper polaritons $|U\rangle
=(|1,ex\rangle -|ph\rangle )/\sqrt{2}$ and dark excitons $|2,ex\rangle $. We
underline that if $\delta _{\epsilon }\neq 0$ both upper polariton
eigenstates $|U-\rangle $ and $|U+\rangle $ contain fractions of the bright
exciton and photon and of the dark exciton. This allows for their direct
optical excitation as well as for the radiative transition between the
states $|U_{1,2}\rangle $ and the lower polariton state $|L\rangle $
accompanied by emission of the THz photon. The rate of the spontaneous
emission of THz radiation can be estimated from the Planck formula:
\begin{equation}
W_{\pm }\approx \frac{\omega ^{3}|d_{\pm }|^{2}n}{3\pi \epsilon _{0}\hbar
c^{3}}F_{P}=W_0F_p  \label{W}
\end{equation}
where $n$ is the refrective index of the cavity, and the dipole
matrix element of the optical transition with emission of a THz
photon in plane of the cavity reads:
\begin{equation}
d_{\pm }=e\langle U\pm |z|L\rangle \approx \frac{16b_{\pm }eL_{z}}{9\pi ^{2}%
\sqrt{2\left( 1+b_{\pm }^{2}\right) }},
\end{equation}
where the last equality holds for a QW of width $L_{z}$ with
infinite barriers. The term $F_{P}$ is the Purcell factor, which
describes the enhancement of the rate of the spontaneous emission of
THz photons due to the presence of the THz cavity. The principal
effect of the cavity is to increase the electric field operator by
the factor of $\sqrt{Q}$\ within the frequency band $\Delta \omega
\approx \omega _{0}/Q$\ around the cavity resonance frequency
$\omega _{0}$, where $Q$ is the quality factor of the cavity
\cite{Gerard}. In the case of a narrow spectral width of the
electronic oscillator coupled to the cavity, this results in the
Purcell formula \cite{Purcell}:
\begin{equation}
F_{p}=\xi Q,
\end{equation}
where $\xi $ is a geometric factor inversely proportional to the
cavity volume $V$ ($\xi \approx 1$ for $V\approx \left( \lambda
/2\right) ^{3}$, where $\lambda $ is the radiation wavelength). In
our case, the electronic resonance is broadened due to short
lifetimes of the initial (UP) and final (LP) states. Applying the
Fermi golden rule to the cavity tuned in resonance with a transition
between homogeneously broadened levels characterized by lifetimes
$\tau _{i\text{ }}$and $\tau _{f}$\ yields the following
generalization of the Purcell formula:
\begin{equation}
F_{p}=\xi \left[ Q^{-1}+\left( \omega _{0}\tau _{i}\right) ^{-1}+\left(
\omega _{0}\tau _{f}\right) ^{-1}\right] ^{-1},
\end{equation}
One can easily find that, for lifetimes of the electronic states of
the order of 10$^{-11}$ s, $F_{p}$ cannot be made much larger than
approximately 50, whatever high is $Q$. This estimate agrees with
experimental results of Ref.\onlinecite{Todorov}.

Generation of THz emission by a microcavity in the polariton lasing regime
is conveniently described by a system of kinetic equations for the upper and
lower polariton states and the THz mode. We consider the following
experimental situation (see Fig.2): the hybridised upper states are
resonantly optically excited. Created in this way polaritons relaxe to the
lower polariton states either directly, emitting THz photons, or via a
cascade of $k\neq 0$ states of the lower polariton mode (acoustic phonon
assisted relaxation). Considering the degenerate states $|U+\rangle $ and $%
|U-\rangle $ as a single upper polariton state $|U\rangle $ \, as
well as all $k\neq0$ states as a single reservoir \cite{Solnyshkov},
the rate equations read

\begin{figure}[tbp]
\begin{center}
\includegraphics[width=0.6\linewidth]{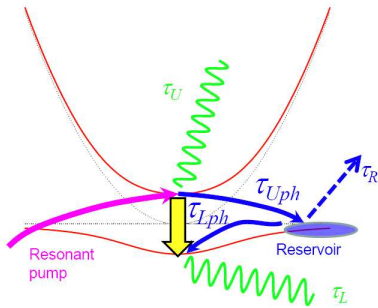}
\end{center}
\caption{The scheme illustrating possible transitions in the system. The
vertical axis is energy, the horisontal axis is the in-plane wave-vector.}
\label{fig2}
\end{figure}

\begin{eqnarray}
\dot{N_{U}} &=&P-\left( \tau _{U}^{-1}+\tau _{UR}^{-1}\right) N_{U}+W\left[
N_{L}N(N_{U}+1)-\right.  \\
&&-\left. N_{U}(N_{L}+1)(N+1)\right] ,  \notag \\
\dot{N_{L}} &=&-\tau _{L}^{-1}N_{L}+\tau _{LR}^{-1}N_{R}+ \\
&&+W\left[ N_{U}(N_{L}+1)(N+1)-N_{L}N(N_{U}+1)\right] ,  \notag \\
\dot{N} &=&-\tau ^{-1}N+W\left[ N_{U}(N_{L}+1)(N+1)-\right.  \\
&&\left. -N_{L}N(N_{U}+1)\right] .  \notag \\
\dot{N_{R}} &=&-\tau _{R}^{-1}N_{R}-\tau _{LR}^{-1}N_{R}(N_{L}+1)+\tau
_{UR}^{-1}N_{U}.
\end{eqnarray}%
where $N_{U},N_{L},N_{R},N_{THz}$ are the populations of the upper polariton
modes, lower polariton mode, the reservoir of $k\neq 0$ states of the lower
polariton mode and the THz photon mode, respectively, $\tau _{U},\tau
_{L},\tau _{R},\tau $ are the lifetimes of these states, $\tau _{UR}$ and $%
\tau _{LR}$ are the rates of acoustic phonon assisted transitions between
the upper polariton mode and the reservoir and between the reservoir and the
lowest energy polariton state, respectively, $P$ is the rate of polariton
generation in the upper mode due to the optical pump, $W$ is a rate of THz
emission given by Eq.\ref{W}. In the stationary regime, the occupation
number of the THz mode can be found from the solution of the above set of
equations putting $\dot{N_{U}}=\dot{N_{L}}=\dot{N}=\dot{N_{R}}=0$.

Figure 3 shows the dependence of the quantum efficiency parameter
$\beta =N/P\tau $ on the pumping rate $P$. The parameters used in
this calculation are: $\tau _{U}=\tau _{L}=20ps,\tau _{R}=100ps,\tau
_{UR}=\tau _{LR}=10ps,\tau/Q =10ps,W/F_{p}=W_{0}=10^{-9}ps^{-1}.$
One can see that for the realistic choice of parameters
corresponding to the existing polariton lasers \cite{Bloch, Levrat}
the quantum efficiency achieves $\beta =1-1.5\%.$
\begin{figure}[tbp]
\begin{center}
\includegraphics[width=0.8\linewidth]{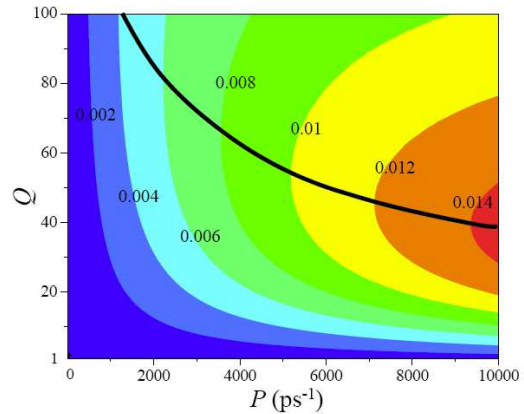}
\end{center}
\caption{Quantum efficiency of the terahertz emitter as a function of the
pump intensity P and a quality factor of the terahertz cavity Q. Black line
shows the dependence of the optimum value of Q on pump intensity.}
\label{fig3}
\end{figure}
Finally we note that the mechanism of THz emission considered here
is qualitatively different from one used in quantum cascade lasers
embedded in THz cavities \cite{Ciuti}. Effectively, here the THz
emission is stimulated by the population of the lowest energy
polariton state, which provides quite high quantum efficiency
compared with the best commercially available quantum cascade
lasers. Semiconductor microcavities in the regime of polariton
lasing may be used as efficient sources of the THz radiation having
a quantum efficiency exceeding 1\% according to our estimations.

This work was supported by EU projects CLERMONT4 and POLALAS, RFBR grant
10-02-01184a. IAS acknowledges the support from RANNIS "Center of excellence
in polaritonics".

\end{document}